\documentclass[epj]{svjour}

\usepackage{graphics}

\begin{document}

\title{Stochastic resonance in bistable confining potentials}
\subtitle{On the role of confinement}

\author{
Els Heinsalu \inst{1,2},
Marco Patriarca \inst{1},
\and Fabio Marchesoni \inst{3}
}

\institute{National Institute of Chemical Physics and Biophysics,
R\"avala 10, 15042 Tallinn, Estonia
\and
IFISC, Instituto de F\'isica Interdisciplinar y Sistemas Complejos (CSIC-UIB),
E-07122 Palma de Mallorca, Spain
\and
Dipartimento di Fisica, Universit\`a di Camerino, I-62032 Camerino, Italy
}

\date{Received: date / Revised version: date}

\abstract{We study the effects of the confining conditions on the
occurrence of stochastic resonance (SR) in continuous bistable
systems. We model such systems by means of double-well potentials
that diverge like $|x|^q$ for $|x|\to \infty$. For super-harmonic
(hard) potentials with $q>2$ the SR peak sharpens with increasing
$q$, whereas for sub-harmonic (soft) potentials, $q<2$, it gets
suppressed.
} 

\maketitle


\section{Introduction}
\label{intro}


The simplest dynamical system displaying stochastic resonance (SR)
is a Brownian particle bound into a one-dimensional double well
under the action of a time oscillating tilt and subjected to
fluctuating forces (noise) \cite{Gammaitoni1998a,Benzi1981a}. The SR
mechanism can be revealed as a maximum in the amplitude of the
periodic component of the average particle position as a function of
the noise intensity (temperature). Due to fluctuations, the particle
randomly jumps between the two potential wells with Kramers rate
\cite{borkovec} that depends on the double well potential and
temperature. When the average escape time of the particle out of the
potential minima (i.e., the inverse of the Kramers rate)
approximately equals the half time-period of the applied
perturbation, the noise induced interwell jumps and the periodic
force synchronize, thus leading to SR.

When studying the problem of a Brownian particle in a symmetric
double well periodically tilted in time, the corresponding potential
$U(x)$ is usually assumed to diverge like $U(x) \sim x^4$ at large
$x$ \cite{Gammaitoni1998a,borkovec}, so as to ensure a robust
confining action. However, the divergence of the potential for $|x|
\to \infty$ strongly affects the response of the system to an
external time-periodic forcing. The goal of the present paper is to
investigate how the Brownian motion in a double well changes with
the confining strength of the one-dimensional potential $U(x)$. For
simplicity we assume that $U(x) \sim |x|^q$ for $x \to \pm \infty$.
By studying the dependence of a SR spectral quantifier on $q$, we
conclude that bistability is a {\it necessary, but not sufficient}
condition for a one-dimensional system to exhibit SR.


\section{Model} \label{model}


The model discussed in the following represents an overdamped
Brownian particle with coordinate $x$. Its dynamics is described by
the Langevin equation,
\begin{equation} \label{L1}
 \eta \dot{x} = - U'(x) + A(t) + \xi (t)\, ,
\end{equation}
where $(\dots)' \equiv \mathrm{d}(\dots)/\mathrm{d}x$. The
confining potential,
\begin{equation}
  \label{potential}
  U(x) = U_0 \exp \left(- x^2/L_0^2 \right) + k |x|^q / q \, ,
\end{equation}
is obtained by superimposing a Gaussian repulsive barrier of height
$U_0$ and width $L_0$, to a power-law potential well. To ensure
confinement, our analysis is restricted to $q >1$. The total
potential is mirror symmetric at $x = 0$, i.e. $U(x) = U(-x)$.
Depending on $q$ a potential $U(x)$ is called hard (super-harmonic) for $q>2$,
or soft (sub-harmonic) for $q<2$ \cite{zannetti}.
The periodic drive $A(t)$ is chosen as
\begin{equation} \label{f-external}
  \label{fext}
 A(t) = A_0 \cos (\Omega t) \, ,
\end{equation}
with amplitude, $A_0$, angular frequency, $\Omega \equiv 2\pi \nu$,
and time origin arbitrarily set to zero. The fluctuating force $\xi
(t)$ is modeled as a stationary zero-mean Gaussian noise with
auto-correlation function $\langle \xi(t) \xi(t') \rangle = 2 \eta
k_\mathrm{B} T \delta (t-t')$. Here $T$ is the temperature and
$\eta$ the friction coefficient.

\begin{figure}
\resizebox{0.75\columnwidth}{!}{%
  \includegraphics{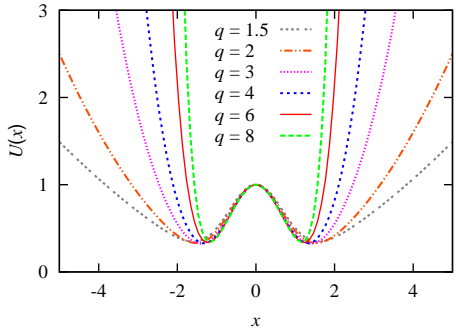}
} \caption{Rescaled potential (\ref{resc-pot}) for $k=0.2$ and $q$
ranging between $1.5$ and $8$. The barrier height is approximately
constant, $\Delta U \simeq 0.66$, and the minima $\pm x_m$ slowly
shrink with $q$ from $x_m\simeq 1.59$ down to $x_m\simeq 1.17$.}
\label{fig_potentials}
\end{figure}

For numerical purposes it is convenient to choose $U_0$, $L_0$, and
$\tau \equiv \eta L_0^2/U_0$ as the new units respectively of
energy, space and time. Correspondingly, the variables and the
parameters appearing in Eq.~(\ref{L1}) can be replaced by the
dimensionless quantities $\tilde{x} = x/L_0$, $\tilde{t} = t/\tau$,
$\tilde{k} = L_0^q k / U_0$, $\tilde{A}_0 = A_0 L_0/U_0$,
$\tilde{\Omega} = \Omega \tau$, and $\tilde{T} = k_B T/U_0$. To avoid
a cumbersome notation, in the following we omit all the tildes. In
dimensionless notation the potential (\ref{potential}) reads,
\begin{equation} \label{resc-pot}
U(x) = \exp(-x^2) + k|x|^q / q \, ,
\end{equation}
and the Langevin equation (\ref{L1}) can be rewritten as
\begin{eqnarray}
  \label{L2}
  \dot x
  = 2 {x} \exp(-x^2) - k |{x}| ^q/x
 + {A}_0 \cos ({\Omega} {t})
  + \sqrt{{T}}\xi ( t) \, ,
\end{eqnarray}
after the Gaussian noise $\xi(t)$ has been further rescaled so that
$\langle \xi({t}) \rangle = 0$ and $\langle \xi({t}) \xi({t}')
\rangle=2 \delta ({t} - {t}')$.
In the following we study how changing $q$ influences the response
of the particle to the periodic forcing signal $A(t)$. As a result
of rescaling, the height, $U_0$, and the width, $L_0$, of the
potential barrier, as well as the friction coefficient, $\eta$, have
been set to one. The remaining tunable parameter $k$ of the
potential (\ref{resc-pot}) will be kept fixed to $k=0.2$ throughout
the present paper. Due to the Gaussian nature of the
potential barrier, the barrier height, $\Delta U$, and the potential
minima, $\pm x_m$, weakly depend on $q$ (see
Fig.~\ref{fig_potentials}); therefore, the observed residual SR
dependence on $q$ is mostly an effect of the varying confining
strength of the potential.

We have simulated the behavior of the system by numerically
integrating the rescaled Langevin equation (\ref{L2}) through a
Milshtein algorithm \cite{Kloeden1999a,Milstein2004a}. Stochastic
trajectories were simulated for different time lengths
$t_\mathrm{max}$ and time steps $\Delta t$, so as to ensure
appropriate numerical accuracy and transient effects subtraction.
Average quantities have been obtained as ensemble averages over at
least $10^4$ trajectories.

\section{Results}
\label{results}


In the long time regime, after transient effects subsided, the
response $\langle x(t) \rangle$ of a particle moving in a symmetric
bistable potential $U(x)$ under the action of the signal
(\ref{f-external}) with small-amplitude, $A_0 x_m \ll \Delta U$, and
low-frequency, $\Omega \ll U''(x_m)$, results from the interplay of
inter- and the intrawell dynamics \cite{Gammaitoni1998a}. On
ignoring for the time being the intrawell dynamics, the system
response at low temperatures is dominated by its harmonic component
\cite{Gammaitoni1998a,McNamara1989a,Presilla1989a,Hu1990a,Jung1991a}
\begin{equation}
  \label{xav}
  \langle x(t \!\to\! \infty) \rangle
  = \bar{x} (T) \cos[\Omega t - \bar{\phi}(T)] \, ,
\end{equation}
with amplitude, $\bar{x}(T)$, and phase, $\bar{\phi}(T)$,
approximated by
\begin{eqnarray}
  \bar{x}(T)&=&
  \frac{A_0 \langle x^2 \rangle _0}{T} \frac{2 r}{\sqrt{4 r^2 + \Omega
  ^2}},
  \label{X0-approx} \\
  \bar{\phi}(T) &=&
  \arctan (\Omega/2 r) \, .
  \label{phi0-approx}
\end{eqnarray}
Here $r \propto \exp(-\Delta U/T)$ is the Kramers rate and $\langle
x^2 \rangle _0$ the variance of the stationary unperturbed process
$x(t)$ ($A_0 = 0$), both temperature dependent quantities. The
amplitude $\bar{x}(T)$ can be manipulated by tuning the noise level.
Note that Eqs. (\ref{xav})-(\ref{phi0-approx}) hold in the linear
response theory limit, only, i.e., for $A_0 x_m \ll T$ and $\Omega >
r$ \cite{Jung1993,schneidman}.

According to Eq.~(\ref{X0-approx}), in the limit $T \to 0$ the
amplitude $\bar{x}(T)$ vanishes due to the potential barrier. The
rate $r$ for the particle to overcome the potential barrier
decreases to zero exponentially when lowering the temperature, that
is $r \ll \Omega$. The interwell jumps are thus inhibited and the
particle gets locked in either minima with probability $1/2$; hence
$\lim_{T\to 0}\langle {x} \rangle =0$. In contrast, for high
temperatures, $T \gg \Delta U$, $r$ may grow much larger than
$\Omega$ and, consequently, $\bar{x}(T) \simeq \langle
x^2\rangle_0/T$. For a hard potential with $q>2$ we show
below that $\langle x^2\rangle_0 \sim T^{2/q}$, so that, again,
$\lim_{T\to \infty} \bar{x}(T)=0$. The occurrence of these limits
for $T\to 0$ and $T\to \infty$ implies the existence of a maximum of
$\bar{x}(T)$ for some optimal $T \sim \Delta U$. This is the
so-called spectral characterization of SR \cite{Gammaitoni1998a}.

\begin{figure}
\resizebox{0.8\columnwidth}{!}{%
\includegraphics{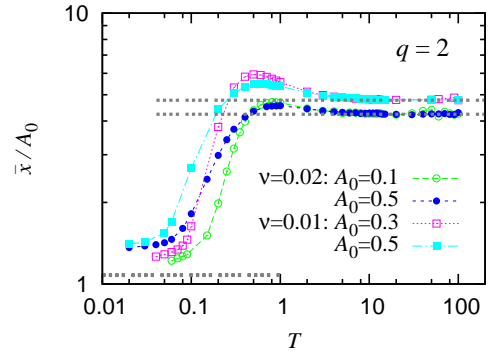}
} \caption{Rescaled amplitude $\bar{x}(T)/A_0$, defined in Eq.~(\ref{xav}),
versus $T$ for the potential~(\ref{resc-pot}) with
$k=0.2$, $q = 2$.
The dashed lines represent the intrawell oscillations, Eq.~(\ref{EQ9}), with $\kappa=
1/|2k\ln(k/2)|$ for $T \to 0$, and $\kappa=k$ in the limit $T \to \infty $.}
\label{fig_amp_Q2}
\end{figure}

\subsection{Harmonic confining potentials}
\label{harmonic_potential}

\begin{figure}
\resizebox{0.74\columnwidth}{!}{%
\includegraphics{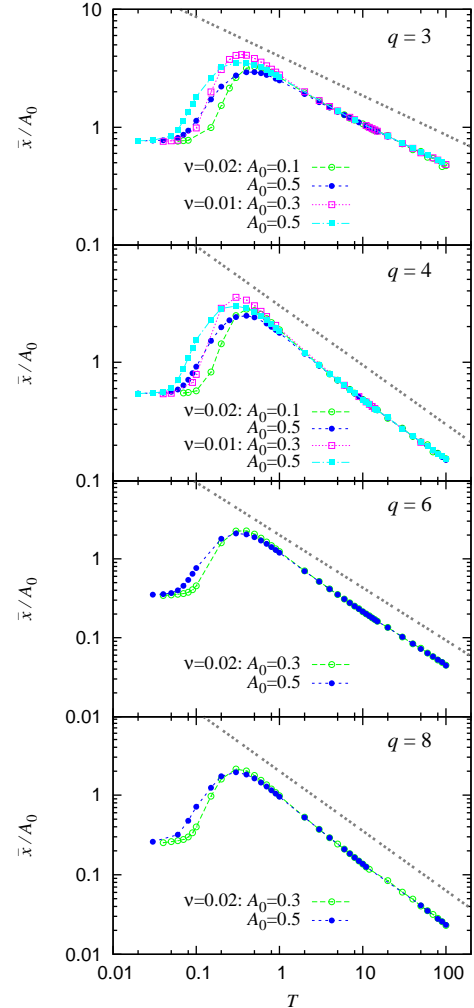}
} \caption{Rescaled amplitude $\bar{x}(T)/A_0$, defined by
Eq.~(\ref{xav}), versus $T$ for the potential (\ref{resc-pot}) with
$k=0.2$ and different $q>2$ (hard potentials). The dashed lines are
the decay power law $T^{2/q-1}$.} \label{fig_amp_Q3-8}
\end{figure}

However, even if the approximate results
(\ref{xav})-(\ref{phi0-approx}) describe correctly the occurrence of
SR in most bistable systems, Figs.~\ref{fig_amp_Q2} and
\ref{fig_amp_Q3-8} ($q>1$) clearly show that for $T \to 0$ the
amplitude $\bar{x}(T)$ approaches a non-zero limit $\bar{x}(0)>0$.
This is a characteristic signature of the intrawell dynamics
\cite{Jung1993,schneidman}. Moreover, for (and only for) $q=2$ a
similar behavior occurs also in the opposite limit $T\to \infty$:
the curves $\bar{x}(T)$ attain an horizontal asymptote, see
Fig.~\ref{fig_amp_Q2}. The coexistence of these two asymptotes,
peculiar to $q=2$, strongly suppresses the SR peak.

The nonzero $\bar{x}(T)$ limits for $T\to 0$ and $T\to \infty$ can
be explained by noticing that an overdamped Brownian particle bound
to a generic harmonic potential well, $U(x)=\kappa (x-x_0)^2/2$,
responds to the signal (\ref{fext}) with amplitude
\begin{equation} \label{EQ9}
\label{x-harmonic} \bar{x} = A_0/\sqrt{\Omega^2 + \kappa^2}.
\end{equation}
[Note also that its variance in the absence of forcing ($A_0=0$) is
$\langle x^2\rangle_0=T/\kappa$.]

In the low temperature limit, $T\to 0$, the particle described by
the Langevin equation (\ref{L2}) is locked in either the right or
left potential well, where it executes additional harmonic
oscillations around the corresponding minima $x_0=\pm x_m$
\cite{Gammaitoni1998a,Jung1993,schneidman,lowD}. Such intrawell
oscillations should not be mistaken for the interwell dynamics
described by Eq.~(\ref{xav}) \cite{Hu1990a}. Their amplitude is well
reproduced by Eq.~(\ref{x-harmonic}) with $\kappa \equiv U''(\pm
x_m)=|2k\ln (k/2)|$.

In the high temperature limit, $T\to \infty$, the fluctuations
$\xi(t)$ may grow so intense that the barrier of the bistable
potential (\ref{resc-pot}) becomes ineffective; the particle is thus
effectively confined into a parabolic potential with $\kappa=k$ and
centered at $x_0=0$. The amplitude of the periodic component of the
particle response to the external force is then described again by
Eq.~(\ref{x-harmonic}) but with $\kappa=k$.

For small frequencies the rescaled amplitude ${\bar x}/A_0$ only depends
on the curvature of the bistable potential at $x_0=\pm x_m$ for
$T\to 0$, ${\bar x}/A_0=1/|2k\ln (k/2)|$, and at $x_0=0$ for $T\to
\infty$, ${\bar x}/A_0=1/k$.

The argument above can be easily generalized for any value of $q$ at
low temperatures, but it becomes untenable in the limit
$T\to \infty$, where nonlinearity comes into play.

\subsection{Hard confining potentials}

As anticipated above, at high temperatures the presence of the
central barrier can be ignored. This implies that for $T\to \infty$
Eq.~(\ref{X0-approx}) simplifies to
\begin{equation} \label{X0-q-1}
\frac{{\bar x}(T)}{A_0} = \frac{\langle x^2 \rangle_0}{T}=
\frac{1}{T}\frac{\int_0^{\infty}dx~ x^2~
\exp{(-kx^q/qT)}}{\int_0^{\infty}dx~ \exp{(-kx^q/qT)}}.
\end{equation}
In Eq.~(\ref{X0-q-1}) we made use of the inequality $r \gg \Omega$
and of the approximate expression $P_0(x) = {\cal N}\exp(-kx^q/qT)$
for the stationary probability density of the unperturbed process
(\ref{L2}); ${\cal N}$ is an appropriate normalization constant.
Note that for sufficiently low $\Omega$, the condition $r \gg
\Omega$ can be consistent with the approximations in
Eq.~(\ref{X0-approx}), whereas suppressing the potential barrier
makes the very definition of $r$ meaningless.

An explicit calculation yields
\begin{equation} \label{X0-q-2}
\frac{{\bar x}(T)}{A_0} =\left( \frac{q}{k}\right)^{2/q}
\frac{\Gamma(3/q)}{\Gamma(1/q)}~\frac{1}{T^{1- 2/q}}.
\end{equation}
Ignoring the algebraic factors we conclude that
\begin{eqnarray}
  \label{X1}
  \lim_{T\to \infty} \bar{x}(T) \sim T^{\, 2/q - 1} \, .
\end{eqnarray}
From here one can see that $\bar{x}$ decreases with increasing $T$ only for hard
confining potentials with $q> 2$. In particular, for the prototypical
case of a quartic potential, $q = 4$ \cite{Gammaitoni1998a}, one
finds  $\bar{x}(T) \sim 1/\sqrt{T}$, as confirmed by the simulation
results (see Fig.~\ref{fig_amp_Q3-8}). For $q = 2$, one
recovers the harmonic limit discussed in the foregoing subsection.

The decay law of $\bar{x}(T)$, Eq.~(\ref{X1}), is clearly a
consequence of the nonlinearity of the potential. Indeed, the same
power law can be recovered by implementing the stochastic
linearization scheme of Ref.~\cite{bulsara}: In Gaussian
approximation, for $q$ an integer, $\lim _{|x|\to \infty}U(x)=\kappa
x^2/2$ with $\kappa = (q-1)!!k\langle x^2 \rangle_0^{q/2-1}$; from
the relation $\langle x^2 \rangle_0=T/\kappa$, holding for
harmonic potentials, Eq.~(\ref{X1}) follows.

Moreover, $\bar{x}(T)$ cannot decrease faster than $T^{-1}$, which
happens for $q \to \infty$. It should be noticed that
$\bar{x}(T)\sim T^{-1}$ is the decay law predicted in two-state
model approximation \cite{McNamara1989a}, where $\langle x^2
\rangle_0$ is replaced by $x_m^2$ (i.e., a constant).

\begin{figure}
\resizebox{0.8\columnwidth}{!}{%
  \includegraphics{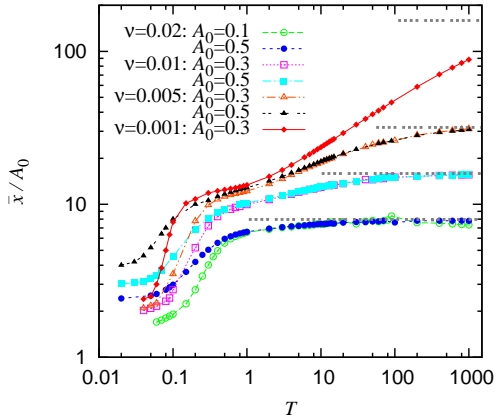}
} \caption{Rescaled amplitude $\bar{x}(T)/A_0$ versus $T$ for the
potential (\ref{resc-pot}) with $k=0.2$ and $q=1.5$ (soft
potential). The dashed lines represent the horizontal asymptotes
$1/\Omega$ (see text). In place of the SR peak an inflexion point is
detectable for low $\Omega=2\pi \nu$.} \label{fig_amp_Q1.5}
\end{figure}

\subsection{Soft confining potentials}

Equation (\ref{X1}) for $q<2$ suggests that $\bar{x}(T)$ may diverge
at high temperatures. However, when dealing with soft potentials,
the linear theory approximations (\ref{xav})-(\ref{phi0-approx})
must be used with caution. In the limit $T\to 0$ the interwell
oscillation amplitude (\ref{X0-approx}) is known to apply only for
very small perturbation amplitudes \cite{zannetti}: This explains
the residual $A_0$ dependence of the low $T$ plateaus reported in
Fig.~\ref{fig_amp_Q1.5}.

More importantly, in the high $T$ limit, although the barrier of a
soft potential is awash with noise, confinement gets so weak that
the particle is driven up and down the potential walls primarily by
the deterministic force $A(t)$, rather than by the noise. [For a
comparison, we remind that a particle falls from $\pm \infty$ down
to $\pm x_m$ in a finite time for $q>2$ and in an infinite time for
$q<2$.] In conclusion, on assuming that the Brownian particle
oscillates as if it were (almost) free, its amplitude would read
\begin{equation}
  \label{X1soft}
  \lim_{T\to \infty} \bar{x}(T) \sim A_0/\Omega \, .
\end{equation}
$\bar{x}(T)$ is then expected to develop high $T$ plateaus also for
$q<2$, but, in contrast with the cases discussed in Sec.~\ref{harmonic_potential},
such plateaus are inverse proportional to
the drive frequency (also for low frequency drives, see
Fig.~\ref{fig_amp_Q1.5}).

In the case of sub-harmonic bistable potentials the hallmark of SR
is thus the monotonic increase of the response amplitude with T, as
opposed to the occurrence of a maximum often detected in the
super-harmonic potentials. Such a behavior resembles the phenomenon
of "SR without tuning" discussed in Ref.~\cite{Collins1995}, with
the important difference that here it has been observed in a {\it
single} unit, rather than in a summing network of $N$ excitable
units.


\section{Conclusions}

We conclude this note with two important remarks:

{\it (i)} The coexistence of two locally stable minima separated by
a potential barrier is commonly advocated to explain the occurrence
of a SR peak in a continuous bistable dynamics. Here we have shown
that this keeps being true as long as the confining action exerted
by the potential is super-harmonic. Most notably, for harmonic and
sub-harmonic potentials the periodic component of the system
response may increase monotonically with the noise level.

{\it (ii)} In many experimental reports (see, for a review, Ref.~\cite{JSP}),
the authors tried to characterize the SR peak by means
of Eq.~(\ref{xav}), without paying much attention to the $T$
dependence of the quantity $\langle x^2 \rangle_0$. In some cases
they adopted an outright two-state model with $\langle x^2
\rangle_0=x_m^2$. This led to a poor fit of the decaying tail of
$\bar{x}(T)$, whereas a more accurate fit could have given a
valuable clue to better model the system at hand \cite{EPR}.



\section*{Acknowledgments}
This work has been supported by the Estonian Science Foundation through
Grant  No. 7466 (M.P., E.H.), Spanish MEC and FEDER through project
FISICOS (FIS2007-60327), and ESF STOCHDYN project (E.H.).


\end{document}